\documentclass[review]{elsarticle}

\usepackage{lineno,hyperref}
\usepackage{color}
\usepackage{amsmath,amsfonts,amssymb,esvect,ulem}
\usepackage{float}
\usepackage{morefloats}
\usepackage{graphicx}
\usepackage{epsfig}
\usepackage{placeins}
\usepackage{esint}
\usepackage{epstopdf}
\usepackage{subcaption}
\usepackage{mathtools}
\modulolinenumbers[5]

\newcommand{\TBF}{\textcolor{red}{To be filled}}

\newcommand{\TAMU}{Texas A\&M University}
\newcommand{\dir}{\hat{\Omega}}
\newcommand{\uvec}{\vec{\Omega}}
\newcommand{\ulength}{\Omega}
\newcommand{\pos}{\vec{r}}

\newcommand{\Sphere}{\mathbb{S}^2}

\newcommand{\SPN}{SP$_N$\ }
\newcommand{\PN}{P$_N$\ }
\newcommand{\Pop}{\mathcal{P}}
\newcommand{\Qop}{\mathcal{Q}}
\newcommand{\Fop}{\mathcal{F}}
\newcommand{\Gop}{\mathcal{G}}
\newcommand{\dt}{\frac{1}{v}\frac{\partial}{\partial t}}

\newcommand{\Hsp}{\mathcal{H}}
\newcommand{\Rsp}{\mathbb{R}^3}
\newcommand{\Fig}{Figure\ }
\newcommand{\Ref}{Reference\ }
\DeclarePairedDelimiter{\ceil}{\lceil}{\rceil}

\journal{\textcolor{red}{\TBF}}

\begin{document}

\begin{frontmatter}

\title{Mathematical and numerical validation of the simplified spherical harmonics approach for time-dependent anisotropic-scattering transport problems in homogeneous media}

\author[mymainaddress]{Can Pu}
\ead{pucan1991@gmail.com}
\author[mymainaddress]{Ryan G. McClarren}
\ead{rgm@tamu.edu}

\address[mymainaddress]{Nuclear Engineering, \TAMU,~College Station, TX 77843-3133}

\begin{abstract}
In this work, we extend the solid harmonics derivation, which was used by Ackroyd et al to derive the steady-state \SPN equations, to transient problems. The derivation expands the angular flux in ordinary surface harmonics but uses harmonic polynomials to generate additional surface spherical harmonic terms to be used in Galerkin projection. The derivation shows the equivalence between the \SPN and the \PN approximation. Also, we use the line source problem and McClarren's ``box" problem to demonstrate such equivalence numerically. Both problems were initially proposed for isotropic scattering, but here we add higher-order scattering moments to them. Results show that the difference between the \SPN and \PN scalar flux solution is at the roundoff level.

\end{abstract}

\begin{keyword}
	Simplified spherical harmonics \sep solid harmonics
\end{keyword}

\end{frontmatter}

\linenumbers

\section{Introduction}
The simplified spherical harmonics (SP$_N$) approximation was initially developed in the 1960s to reduce the number of degrees of freedom required to solve transport problems in multiple dimensions using moment-based methods such as the spherical harmonics (P$_N$) expansion \cite{Gelbard1960:SPn,Gelbard1961:SPn,Gelbard1962:SPn}. Initially, the \SPN approximation was ``derived" by Gelbard in an ad-hoc way by manipulating the P$_N$ expansion from 1-D slab geometry to have a 3-D form.
The theoretical study of the \SPN equations underwent a great awakening in the 1990s. Larsen, Morel and McGhee \cite{Larsen1993:SPn} made an asymptotic derivation of the \SPN equations and showed that the accuracy of the approximation is dependent upon the scattering ratio of the system. Subsequent works extended the asymptotic analysis to anisotropic or time-dependent problems \cite{Larsen1996:SPn,Tomavsevic1996:SP2,Brantley2000:SP3,Frank2007:SPn,Olbrant2013:SPn}. However, the use of a Neumann series expansion to the unbounded streaming operator makes the asymptotic analysis a formal proof.

Another way to the \SPN equations follows the signposts of the solid harmonics. By generalizing the angular domain from $\Sphere$ (i.e., the unit sphere) to $\Rsp$ Ackroyd et al. used properties of solid harmonics to introduce additional harmonics, perform Galerkin projection, restrict the relaxed equations back to $\Sphere$, and thereby derive the \SPN equations for steady-state, anisotropic scattering problems in homogeneous media \cite{Ackroyd1999:IsotropicSHPn,Ackroyd1999:AnisotropicSHPn}. Later, Hanu\v{s} used solid harmonics in tensor form and derived the SP$_3$ equations for steady-state isotropic-scattering problems \cite{Hanus2014:MCPn}; Chao was also able to use the solid harmonics approach to show that Pomraning's angular flux trial space, which can lead to the \SPN equations \cite{Pomraning1993:SPn}, is a particular solution to the steady-state isotropic-scattering transport equation \cite{Chao2016:SPn}.

In this contribution, we would extend the solid harmonics derivation to time-dependent anisotropic-scattering homogeneous-media problems and numerically demonstrate the equivalence between \SPN and \PN equations. As such, we extend, albeit incrementally, the domain of applicability of the \SPN\ equations to time-dependent, homogeneous problems with arbitrary scattering, a result that does not appear in the extant literature.

\section{Solid Harmonics for Time-Dependent Transport Problems}
As introduced in the previous section, the derivation below is based on solid harmonics which were also used by \Ref\cite{Ackroyd1999:IsotropicSHPn} and \cite{Ackroyd1999:AnisotropicSHPn}. Some of the properties of the solid harmonics that we use can be found in the papers, but for clarity in exposition are repeated herein.

\subsection{Expansion of the Transport Equation}\label{SPnDerivationSection:TheGroupedPnExpansionOfTransportEquation}
For monoenergetic neutron transport in homogeneous media, we begin with the equation \cite{Bell_Glasstone,2007PhDT}
\begin{equation}
    \label{TransportEquation}
    \dt\psi(\pos,\dir,t)+\dir\cdot\nabla\psi(\pos,\dir,t)+\sigma_t\psi(\pos,\dir,t)=\oint_{\Sphere}\psi(\pos,\dir',t)\sigma_s(\dir\cdot\dir')d\Omega'+q(\pos,\dir,t),
\end{equation}
where the integration over $\Sphere$ is often written as an integral over $4\pi$ in the transport literature.

The \PN approximation expands the angular flux and the anisotropic external source in terms of surface spherical harmonics
\begin{align}
    \psi(\pos,\dir,t)&=\sum_{l=0}^\infty\frac{2l+1}{4\pi}\psi_l(\pos,\dir,t),\\
    q(\pos,\dir,t)&=\sum_{l=0}^\infty\frac{2l+1}{4\pi}q_l(\pos,\dir,t),
\end{align}
where
\begin{align}
    \psi_l(\pos,\dir,t)&=\sum_{m=-l}^l\psi_l^m(\pos,t)Y_l^m(\dir),\\
    q_l(\pos,\dir,t)&=\sum_{m=-l}^lq_l^m(\pos,t)Y_l^m(\dir),
\end{align}
and $Y_l^m$ are the surface spherical harmonics \cite{Bell_Glasstone}, commonly referred to without the ``surface'' adjective.

The scattering cross-section can be expanded into
\begin{equation}
    \sigma_s(\dir\cdot\dir')=\sum_{l=0}^\infty\frac{2l+1}{4\pi}\sigma_{sl}\sum_{m=-l}^lY_l^m(\dir)Y_l^m(\dir'),
\end{equation}
so that, by defining
\begin{equation}
    \sigma_l:=\sigma_t-\sigma_{sl},
\end{equation}
Equation (\ref{TransportEquation}), by orthogonality of surface spherical harmonics, is written as the sum over $l$-moments
\begin{equation}
    \label{ExpandedTransportEquation}
    \sum_{l=0}^\infty(2l+1)\left[\left(\dt+\dir\cdot\nabla+\sigma_l\right)\psi_l(\pos,\dir,t)-q_l(\pos,\dir,t)\right]=0.
\end{equation}

\subsection{Galerkin Projection}\label{SPnDerivationSection:GalerkinProjection}
The next step is to split the streaming term $(2l+1)\dir\cdot\nabla\psi_l(\pos,\dir,t)$ into harmonics of degree $l-1$ and $l+1$ respectively, and this is where solid harmonics come to play their role.

Harmonic polynomials, often referred to as solid harmonics, are polynomials that are solutions to the Laplace equation, and a surface spherical harmonic of degree $l$ is the restriction onto $\Sphere$ of a homogeneous harmonic polynomial on $\mathbb{R}^3$ of degree $l$ whose collection is denoted as $\Hsp_l(\Rsp) $\cite{Axler2013:HarmonicFunctionTheory}. For $\uvec\in\mathbb{R}^3$, solid harmonics can be obtained by rescaling surface harmonics by a factor of the length of $\uvec$ to the $l$th power, $\ulength^l$, and the process relaxes the restriction of the domain from $\Sphere$ to $\mathbb{R}^3$. In other words, if $Y_l^m(\dir)$ is a surface harmonic, then $\ulength^l Y_l^m(\uvec)$ is a solid harmonic. For convenience, when $p_l$ is a surface harmonic, we denote the associated solid harmonic as $\tilde{p}_l$.

Now we make the following definitions for $\uvec\in\mathbb{R}^3$
\begin{equation}
    \nabla_\ulength:=\left(\frac{\partial}{\partial\ulength_x},\frac{\partial}{\partial\ulength_y},\frac{\partial}{\partial \ulength_z}\right)^\text{T},
\end{equation}
\begin{equation}
    \Pop:=\nabla_\ulength\cdot\nabla,\qquad
    \Qop:=\uvec\cdot\nabla,
\end{equation}
and rewrite the streaming term $(2l+1)\dir\cdot\nabla\psi_l(\pos,\dir,t)$ as
\begin{equation}
    \label{SplittingOfStreamingTerm}
    (2l+1)\dir\cdot\nabla\psi_l(\pos,\dir,t)=\left[(2l+1)\Qop\tilde{\psi}_l-\ulength^2\Pop\tilde{\psi}_l+\Pop\tilde{\psi}_l\right]_{\ulength=1}(\pos,\dir,t),
\end{equation}
where the subscript $\ulength=1$ denotes that the term inside brackets is restricted to the unit sphere.

It is proved in Reference \cite{Ackroyd1999:IsotropicSHPn} and shown in \ref{Appendix:ProofOfTwoHarmonicPolynomials} that the term $\left[(2l+1)\Qop\tilde{\psi}_l-\ulength^2\Pop\tilde{\psi}_l\right]$ is a solid harmonic of degree $l+1$ and $\Pop\tilde{\psi}_l$ is a solid harmonic of degree $l-1$. Ergo, when restricted back to $\Sphere$, we get surface harmonics of degree $l-1$ and $l+1$.

Therefore, with Galerkin projection applied to Equation (\ref{ExpandedTransportEquation}), we get
\begin{equation}
    \label{PrimitiveProjectedGroupedTransportEquation}
    \left[(2n+1)\left(\dt+\sigma_n\right)\tilde{\psi}_n+\left((2n-1)\Qop-\ulength^2\Pop\right)\tilde{\psi}_{n-1}+\Pop\tilde{\psi}_{n+1}-(2n+1)\tilde{q}_n\right]_{\ulength=1}=0.
\end{equation}
To proceed, we will define two more operators:
\begin{equation}
    \Fop\psi_n(\pos,\dir,t):=\left[\left((2n+1)\Qop-\ulength^2\Pop\right)\tilde{\psi}_n\right]_{\ulength=1}(\pos,\dir,t),
\end{equation}
and
\begin{equation}\label{eq:Gn}
    \Gop\psi_n(\pos,\dir,t):=\left[\Pop\tilde{\psi}_n\right]_{\ulength=1}(\pos,\dir,t).
\end{equation}
With these definitions Eq.~(\ref{PrimitiveProjectedGroupedTransportEquation}) becomes
\begin{equation}
    \label{GalerkinProjectedGroupedTransportEquation}
    (2n+1)\left(\dt+\sigma_n\right)\psi_n(\pos,\dir,t)+\Fop\psi_{n-1}(\pos,\dir,t)+\Gop\psi_{n+1}(\pos,\dir,t)=(2n+1)q_n(\pos,\dir,t).
\end{equation}
It is this form of the moment equations that we will manipulate to derive time-dependent \SPN equations.

\subsection{Derivation of the \SPN Equations}\label{SPnDerivationSection:TheLastStep}
To further simplify Equation (\ref{GalerkinProjectedGroupedTransportEquation}), two useful identities from Reference \cite{Ackroyd1999:IsotropicSHPn} would be used
\begin{equation}
    \label{IntermediateIdentity(1)}
    \left(\Pop\Qop-\Qop\Pop-\nabla^2\right)\tilde{\psi}_n=0,
\end{equation}
\begin{equation}
    \label{IntermediateIdentity(2)}
    \Pop\ulength^2\tilde{\psi}_n=(\ulength^2\Pop+2\Qop)\tilde{\psi}_n.
\end{equation}
The two identities can be demonstrated using the definition of the underlying operators. Together, Eqs.~\eqref{IntermediateIdentity(1)} and \eqref{IntermediateIdentity(2)}  imply
\begin{equation}
    \label{DirectIntermediateIdentity}
    \Pop^{n+1}\left[(2n+1)\Qop-\ulength^2\Pop\right]\tilde{\psi}_n=(n+1)^2\nabla^2\Pop^n\tilde{\psi}_n,
\end{equation}
as shown in \ref{Appendix:ProofOfDirectIntermediateIdentity}. From these we can use the definition in Eq.~\eqref{eq:Gn} to obtain
\begin{equation}
    \Gop^n\Fop\psi_{n-1}=n^2\nabla^2\Gop^{n-1}\psi_{n-1}.
\end{equation}
Therefore, applying $\Gop$ $n$ times to Equation (\ref{GalerkinProjectedGroupedTransportEquation}) yields
\begin{equation}
    \label{SPnIntermediateSubstitution(1)}
    (2n+1)\dt\Gop^n\psi_n+(2n+1)\sigma_n\Gop^n\psi_n+n^2\nabla^2\Gop^{n-1}\psi_{n-1}+\Gop^{n+1}\psi_{n+1}=(2n+1)\Gop^nq_n.
\end{equation}
Notice that $\Gop^n\psi_n$ and $\Gop^nq_n$ are surface harmonics of degree 0 and thus have no angular dependence, and that the inverse Laplace operator is well-defined in an infinite-medium problem. Using these facts, we can make the following substitution
\begin{equation}
    F_n(\pos)=\frac{1}{n!}\nabla^{-2\ceil{\frac{n}{2}}}\Gop^n\psi_n,
\end{equation}
\begin{equation}
    G_n(\pos)=\frac{1}{n!}\nabla^{-2\ceil{\frac{n}{2}}}\Gop^nq_n.
\end{equation}
To these definitions we can relate the \SPN unknowns as, for $n$ a non-negative integer,
\begin{equation}
    \label{SPnMomentDefinition}
        \phi_{2n}=F_{2n},\qquad
        \vec{\phi}_{2n+1}=\nabla F_{2n+1},
\end{equation}
and write the source as
\begin{equation}
        Q_{2n}=G_{2n}, \qquad
        \vec{Q}_{2n+1}=\nabla G_{2n+1}.
\end{equation}
These definitions will
eventually lead us to the \SPN equations
\begin{equation}
    \label{SPnEquations:EvenEquation}
    \dt\phi_n+\sigma_n\phi_n+\frac{n}{2n+1}\nabla\cdot\vec{\phi}_{n-1}+\frac{n+1}{2n+1}\nabla\cdot\vec{\phi}_{n+1}=Q_n\ \ \ (n\text{  even}),
\end{equation}
\begin{equation}
    \label{SPnEquations:OddEquation}
    \dt\vec{\phi}_n+\sigma_n\vec{\phi}_n+\frac{n}{2n+1}\nabla\phi_{n-1}+\frac{n+1}{2n+1}\nabla\phi_{n+1}=\vec{Q}_n\ \ \ (n\text{  odd}).
\end{equation}

Therefore, we have started with the time-dependent transport equation in a homogeneous medium and demonstrated that we can derive the \SPN equations using properties of the solid harmonics. Moreover, due to the way  the derivation proceeded, a truncated \PN expansion should be equivalent to an \SPN solution truncated at the same level. Therefore, we expect that in a general problem, with homogeneous media, there to be an equivalence between the \SPN and \PN solutions.

\section{Numerical Examples}
In this section, we perform 2D simulations to demonstrate the equivalence between the \SPN and the \PN approximation for time-dependent, anisotropic scattering problems.

We solve the time-dependent \SPN and \PN equations under the implicit Euler scheme, so the problem is essentially a sequence of steady-state problems; the finite difference method is adopted so that all first-order derivatives are approximated by central differentiation; all moments are nodal values evaluated at grid points.
For all the numerical examples below, periodic boundary conditions are enforced, the external source is isotropic and scattering is anisotropic.

\subsection{An Anisotropic Box Problem}
The first test problem comes from Reference \cite{McClarren2010:SPnReview}, where McClarren used it to demonstrate the equivalence between steady-state \PN and \SPN, and Reference \cite{Seibold2014:StaRMAP} refers to it as the ``box" problem. The original problem has $\sigma_t=1\ cm^{-1}$ and isotropic scattering $\sigma_{s0}=0.1\ cm^{-1}$; here we add additional anisotropic scattering $\sigma_{s1}=0.1\ cm^{-1}=\sigma_{s2}=0.1\ cm^{-1}$ to show that such equivalence still holds for time-dependent anisotropic-scattering problems. The external source remains prescribed by
\begin{equation}
    Q(x,y)=
    \begin{cases}
    1   &   1.75\leq x\leq 2.25,1.75\leq y\leq 2.25\\
    1   &   2.75\leq x\leq 3.25,1.5\leq y\leq 2.5\\
    1   &   1.75\leq x\leq 2.25,2.75\leq y\leq 3.25\\
    1   &   3.5\leq x\leq 4.25,3.5\leq y\leq 3.75\\
    0   &   \text{otherwise}
    \end{cases}
\end{equation}

\begin{figure}[!htb]
\center
\begin{minipage}{0.49\linewidth}
\center
\includegraphics[width=5cm]{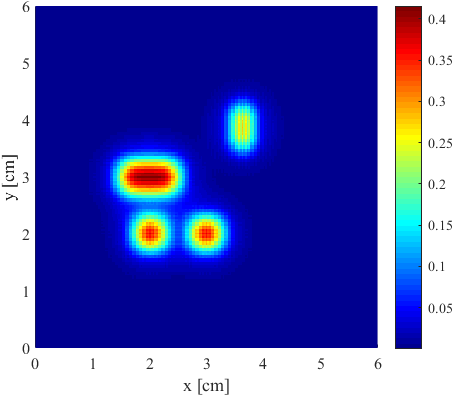}
\subcaption{$T=0.75\ \text{s}$}
\end{minipage}
\begin{minipage}{0.49\linewidth}
\center
\includegraphics[width=5cm]{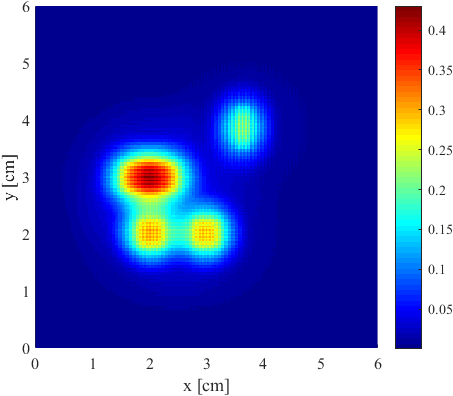}
\subcaption{$T=1.5\ \text{s}$}
\end{minipage}
\caption{Comparison of SP$_3$ scalar flux at $T=0.75\ \text{s}$ and at $T=1.5\ \text{s}$}
\label{AnisotropicBoxProblem:SP3AtT=0.75AndT=1.5}
\end{figure}

We perform the computation with $\Delta x=0.03\ \text{cm}$ and $\Delta t=0.03\ \text{s}$. The SP$_3$ scalar flux at $T=0.75\ \text{s}$ is compared with that at $T=1.5\ \text{s}$ in \Fig\ref{AnisotropicBoxProblem:SP3AtT=0.75AndT=1.5}; it can be seen that at $T=0.75\ \text{s}$, the problem is still in a transient state, thus we choose it as the final time of simulation.

\begin{figure}[!htb]
\center
\begin{minipage}{0.49\linewidth}
\center
\includegraphics[width=5cm]{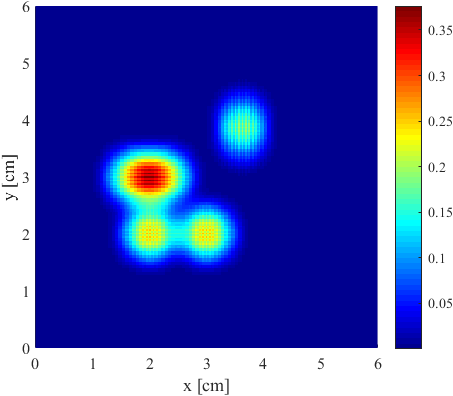}
\subcaption{SP$_1$}
\end{minipage}
\begin{minipage}{0.49\linewidth}
\center
\includegraphics[width=5cm]{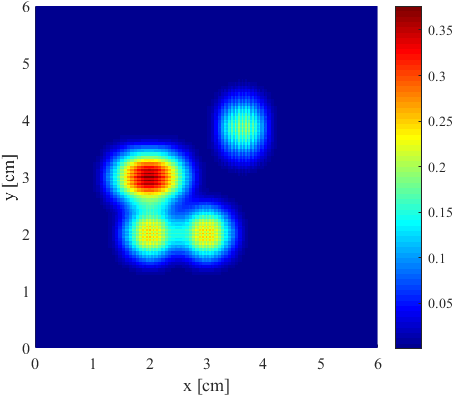}
\subcaption{P$_1$}
\end{minipage}
\begin{minipage}{0.49\linewidth}
\center
\includegraphics[width=5cm]{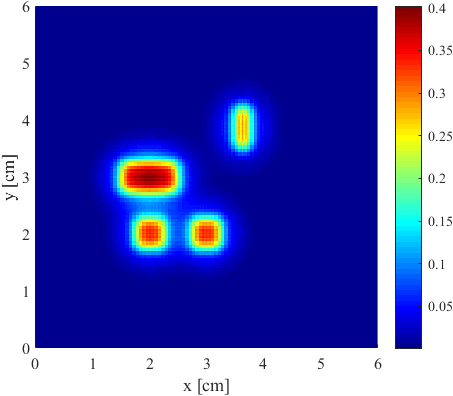}
\subcaption{SP$_5$}
\end{minipage}
\begin{minipage}{0.49\linewidth}
\center
\includegraphics[width=5cm]{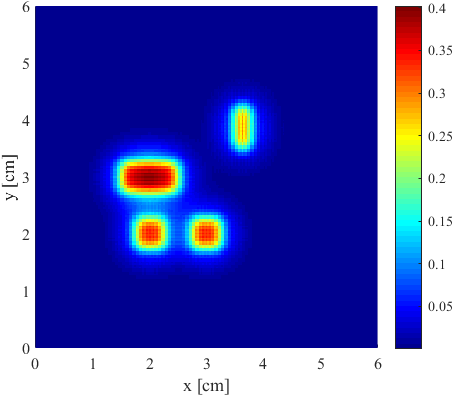}
\subcaption{P$_5$}
\end{minipage}
\caption{Scalar flux computed at $T=0.75\ \text{s}$ for $N=1$ and $N=5$}
\label{BoxProblem:2dPlotN=1And5}
\end{figure}

\begin{figure}[!htb]
\center
\begin{minipage}{0.49\linewidth}
\center
\includegraphics[width=5cm]{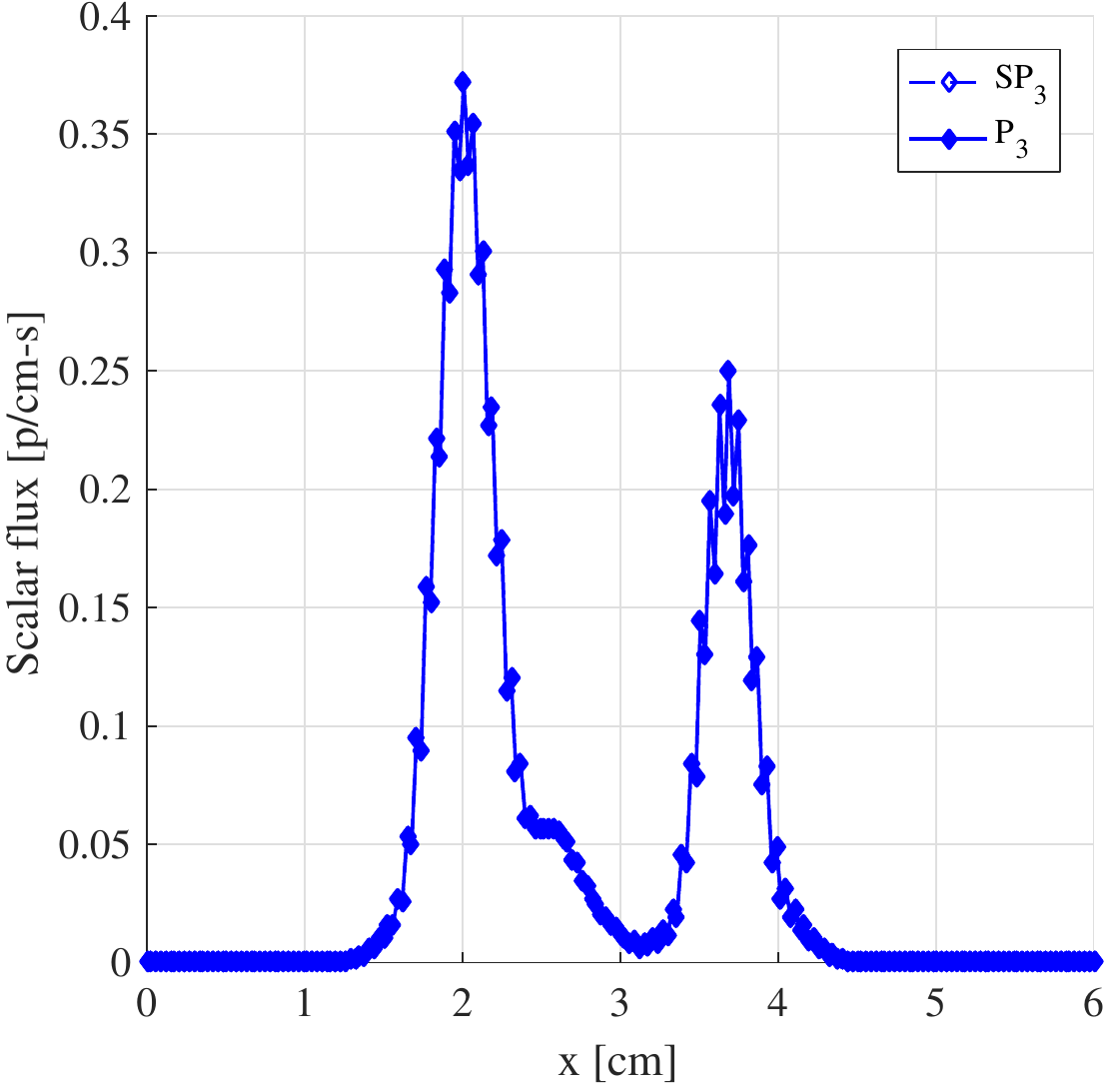}
\subcaption{$N=3$}
\end{minipage}
\begin{minipage}{0.49\linewidth}
\center
\includegraphics[width=5cm]{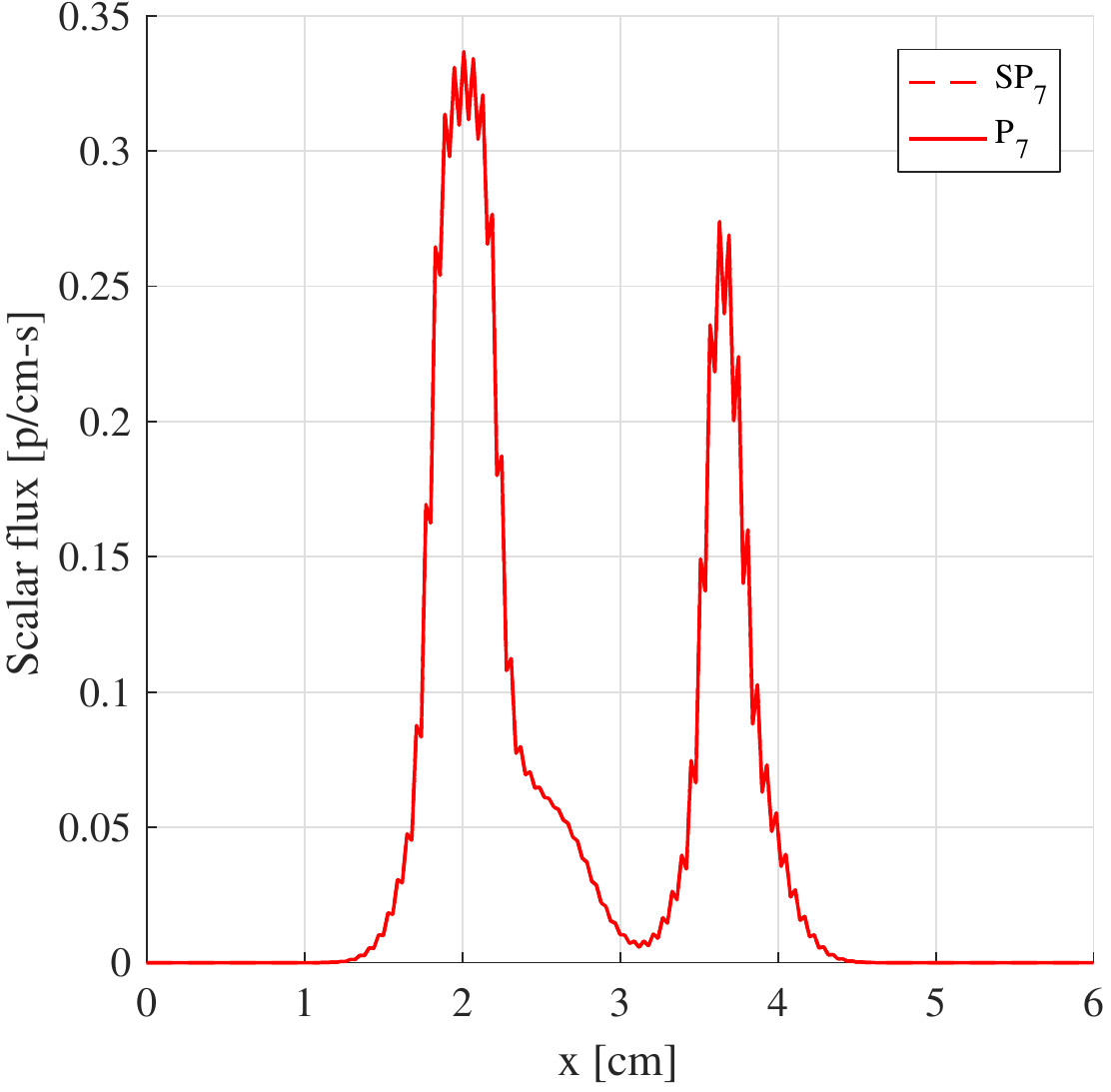}
\subcaption{$N=7$}
\end{minipage}
\caption{Scalar flux computed at $T=0.75\ \text{s}$ along $y=x$ for $N=3$ and $N=7$}
\label{BoxProblem:DiagonalPlotN=3And7}
\end{figure}

The scalar flux solutions for $N=1$ and $N=5$ at $T=0.75\ \text{s}$ are shown in \Fig\ref{BoxProblem:2dPlotN=1And5}; we also plot the scalar flux along $y=x$ for $N=3$ and $N=7$ in \Fig\ref{BoxProblem:DiagonalPlotN=3And7}.

As can be seen from \Fig\ref{BoxProblem:2dPlotN=1And5}-\ref{BoxProblem:DiagonalPlotN=3And7}, despite the existence of oscillations, the \SPN solutions agree with the \PN solutions. In fact, with $N$ ranging from 1 to 7, the $L^\infty$ norm of the deviation from \SPN solution to \PN solution does not exceed $1.6653\times10^{-16}$, so we can claim that the two methods generate identical results.

\subsection{An Anisotropic Line Source Problem}
The line source (pulse) problem is a typical example to demonstrate the Gibbs phenomenon encountered by spectral methods, and it is also important as it is the Green's function for problems with isotropic scattering and source. Here we modify the problem configuration to make the scattering anisotropic such that $\sigma_t=\sigma_{s0}=1\ cm^{-1}$, $\sigma_{s1}=0.5\ cm^{-1}$, $\sigma_{s2}=0.25\ cm^{-1}$ and $\sigma_{s3}=0.125\ cm^{-1}$, but the initial angular flux remains isotropic. To avoid the the singularity in the initial condition, the initial scalar flux $\phi(\vec{r},0)$, as suggested by Reference \cite{Seibold2014:StaRMAP}, is replaced by the distribution:
\begin{equation}
    \label{TestProblem1:InitialScalarFlux}
    \phi(\vec{r},0)=\frac{1}{4\pi s}\exp\left(-\frac{r^2}{4s}\right)
\end{equation}
where $s=3.2\times10^{-4}$.

\begin{figure}[!htb]
\center
\includegraphics[width=5cm]{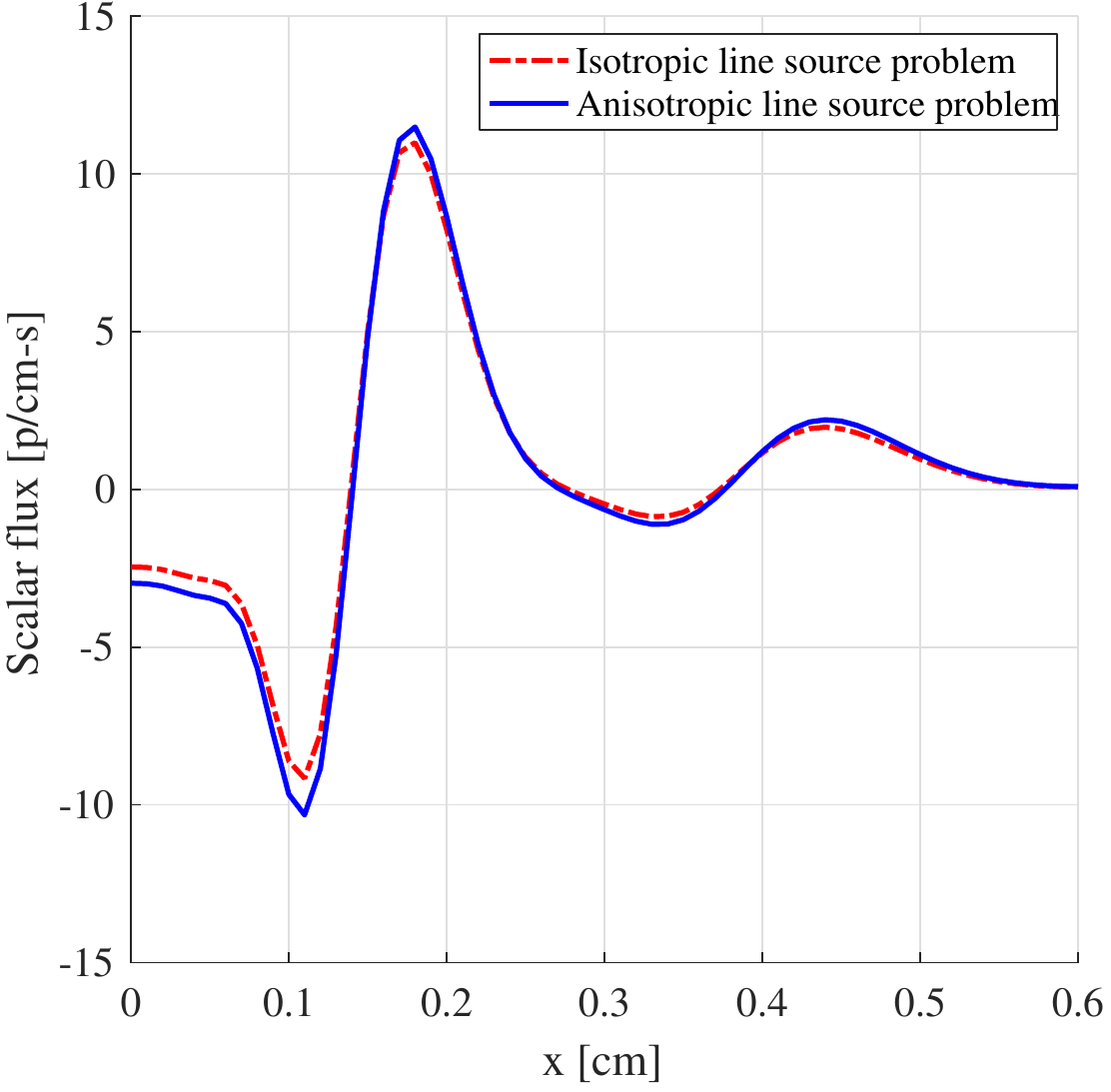}
\caption{Comparison of SP$_3$ scalar flux of the original and the isotropic line source problem on a cut along the positive $x$-axis at $T=0.5\ \text{s}$}
\label{AnisotropicLineSourceProblem:CompareSP3WithIsotropicProblemCutAlongY=0AtT=0.5}
\end{figure}

We perform numerical computation for $T=0.5\ \text{s}$ with $\Delta x=0.01\ \text{cm}$ and $\Delta t=0.005\ \text{s}$. \Fig\ref{AnisotropicLineSourceProblem:CompareSP3WithIsotropicProblemCutAlongY=0AtT=0.5} shows that the solution of the anisotropic line source problem is different from that of the original line source problem, where the scattering is isotropic. The $L^\infty$ norm of the difference is 1.1810.

\begin{figure}[!htb]
\center
\begin{minipage}{0.49\linewidth}
\center
\includegraphics[width=5cm]{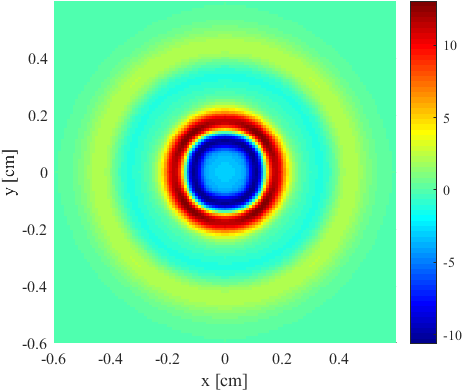}
\subcaption{SP$_3$}
\end{minipage}
\begin{minipage}{0.49\linewidth}
\center
\includegraphics[width=5cm]{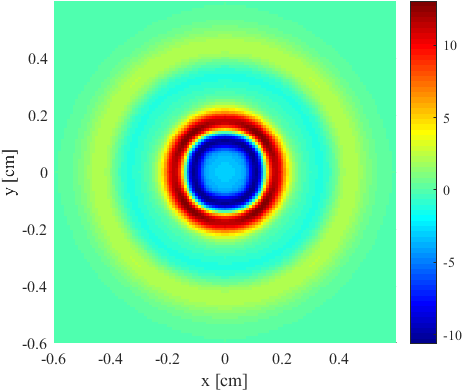}
\subcaption{P$_3$}
\end{minipage}
\begin{minipage}{0.49\linewidth}
\center
\includegraphics[width=5cm]{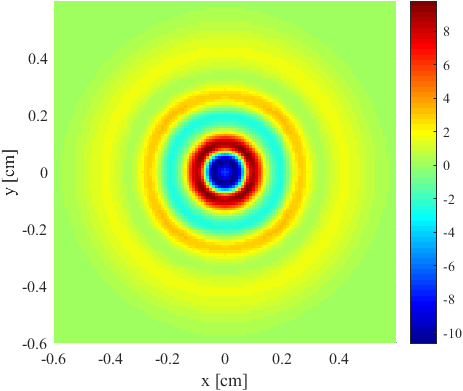}
\subcaption{SP$_7$}
\end{minipage}
\begin{minipage}{0.49\linewidth}
\center
\includegraphics[width=5cm]{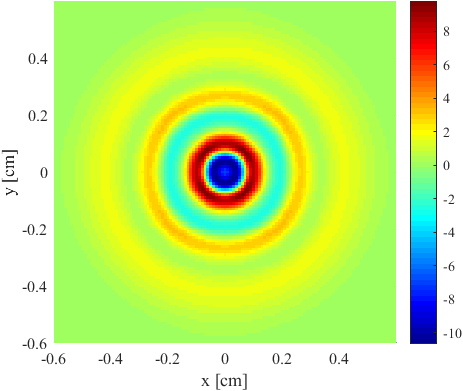}
\subcaption{P$_7$}
\end{minipage}
\caption{Scalar flux for $N=3$ and 7 at $T=0.5\ \text{s}$}
\label{AnisotropicLineSourceProblem:Order3And7AtT=0.5}
\end{figure}

The scalar flux solutions for $N=3$ and $N=7$ are compared in \Fig\ref{AnisotropicLineSourceProblem:Order3And7AtT=0.5}, where one can easily tell that the \SPN and \PN solutions are almost identical.

\begin{figure}[!htb]
\center
\includegraphics[width=6cm]{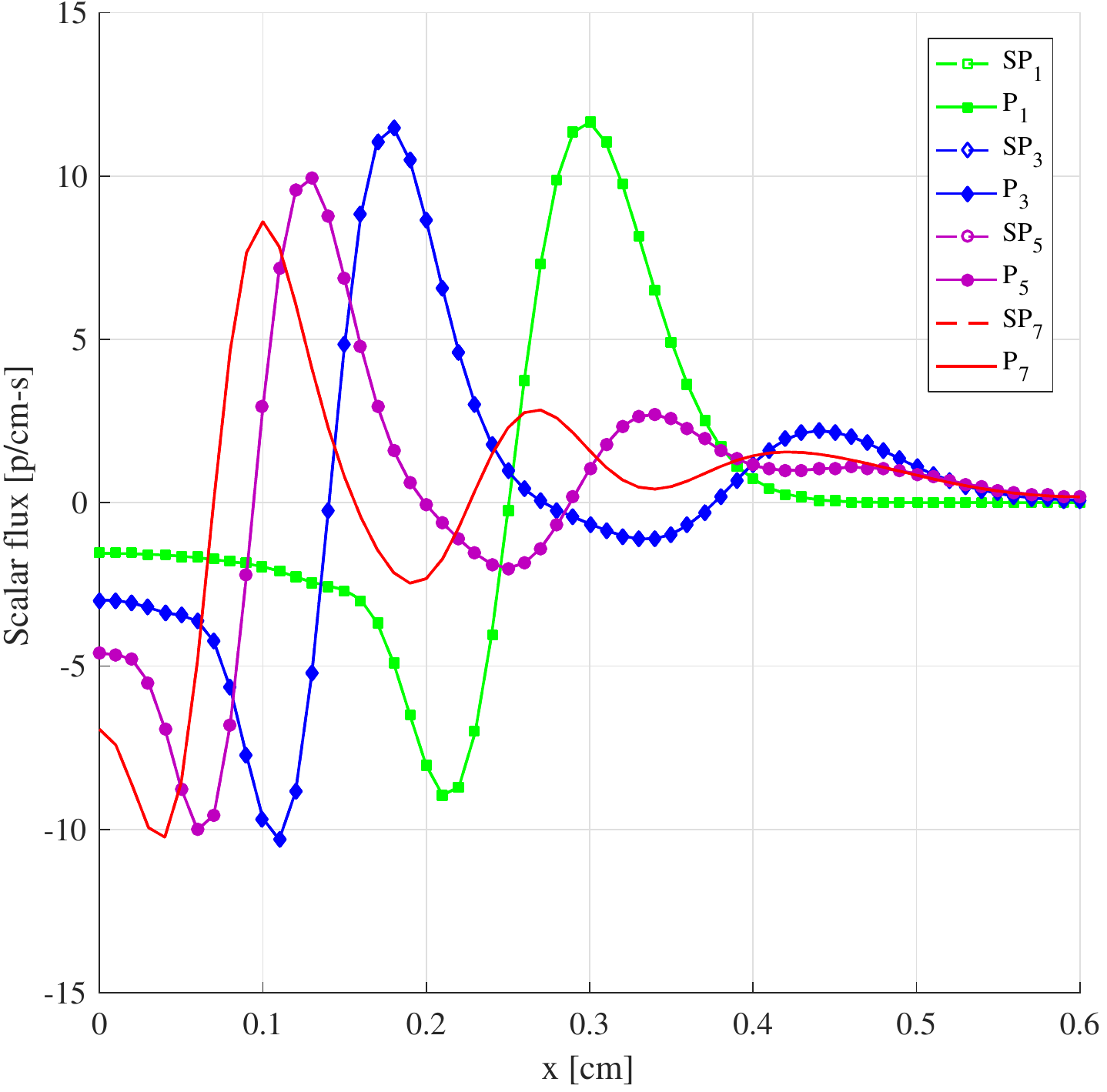}
\caption{Scalar flux on a cut along the positive $x$-axis at $T=0.5\ \text{s}$}
\label{AnisotropicLineSourceProblem:CutAlongY=0AtT=0.5}
\end{figure}

We also plot the scalar flux along the positive $x$-axis in \Fig\ref{AnisotropicLineSourceProblem:CutAlongY=0AtT=0.5} for odd $N$ from 1 to 7. Though not shown in these figures, the \SPN solutions also agree with the \PN solutions for $N$=2, 4 and 6. For $N$ ranging from 1 to 7, the maximum of the $L^\infty$ norms of the \SPN and \PN scalar flux difference is $1.9540\times10^{-14}$.

\FloatBarrier

\section{Discussions and Conclusions}
With the use of harmonic polynomials, we derived the \SPN equations starting from the monoenergetic transport equation in homogeneous media. The derivation implies that we can expect the same scalar flux solution from either the \SPN or \PN approximation.

The theory is then followed by two numerical examples which are modified from the box problem and the line source problem to have anisotropic scattering. In both examples, the \SPN and \PN solutions match very well as the theoretical prediction.

We hope that the work can extend the research and application of the \SPN method. In fact, the study of the \SPN equations is far from over because the strongest theoretical results exist for homogenous, infinite media. The boundary and interface conditions the give SP$_N$-P$_N$ equivalence in heterogeneous problems are a major, open problem. A less ambitious, but useful result, would be the development an algorithm that efficiently generates \SPN moments for anisotropic external source.

\section*{Acknowledgement}
This project is funded, in part, by Department of Energy NEUP research grant from
Battelle Energy Alliance, LLC- Idaho National Laboratory, Contract No: C12-00281.

\bibliography{mybibfile}

\begin{appendix}
\section{Proof that the Two Terms in Equation (\ref{SplittingOfStreamingTerm}) are Solid Harmonics}\label{Appendix:ProofOfTwoHarmonicPolynomials}
We will show how to prove that, if $p_l\in\Hsp_l(\Rsp)$, then $\Pop p_n\in\Hsp_{l-1}(\Rsp)$ and $\left[(2l+1)\Qop-\ulength^2\Pop\right]p_l\in\Hsp_{l+1}(\Rsp)$. The proof here is adopted from Reference \cite{Chao2016:SPn} and Reference \cite{Ackroyd1999:IsotropicSHPn} has a different way. To prove that a function is a solid harmonic of degree $l$, it would be sufficient to show that it is a homogeneous polynomial of degree $l$ and that it is a solution to the Laplace equation.

First, let us consider the easy one, $\Pop p_n$. It is apparent that $\nabla_\ulength$ would result in a vector whose components are homogeneous polynomials of degree $l-1$, so $\nabla\cdot\nabla_\ulength p_l$ satisfies the first condition. To prove that $\nabla_\ulength^2\nabla\cdot\nabla_\ulength p_l=0$, one simply needs to commute the two operators and use the property that $\nabla_\ulength^2 p_l=0$
\begin{equation}
    \nabla_\ulength^2\nabla\cdot\nabla_\ulength p_l
    =\nabla\cdot\nabla_\ulength\nabla_\ulength^2 p_l
    =0
\end{equation}
So we have confirmed that $\Pop p_l\in\Hsp_{l-1}(\Rsp)$.

Proving $\left[(2l+1)\Qop-\ulength^2\Pop\right]p_l\in\Hsp_{l+1}(\Rsp)$ takes slightly more work. Since both $\Qop p_l$ and $\Omega^2\Pop p_l$ are homogeneous polynomials of degree $l+1$, the first condition is again easily met. The next is to prove
\begin{equation}
    \label{SecondConditionForTheHigherOrderSolidHarmonic}
    \nabla_\ulength^2\left((2l+1)\Qop\right)p_l=\nabla_\ulength^2\ulength^2\Pop p_l
\end{equation}
The way is to first prove, via simple vector algebra, that
\begin{equation}
    \label{SecondConditionForTheHigherOrderSolidHarmonic:IntermediateEquation1}
    \nabla_\ulength^2\ulength^2\Pop p_l=(6+4\uvec\cdot\nabla_\ulength+\ulength^2\nabla_\ulength^2)\Pop p_l=(6+4\uvec\cdot\nabla_\ulength)\Pop p_l
\end{equation}
and that
\begin{equation}
    \label{SecondConditionForTheHigherOrderSolidHarmonic:IntermediateEquation2}
    \nabla_\ulength^2\Qop p_l=(2\Pop+\Qop\nabla_\ulength^2)p_l=2\Pop p_l
\end{equation}
By Euler's homogeneous function theorem
\begin{equation}
    \uvec\cdot\nabla_\ulength\Pop p_l=(l-1)\Pop p_l
\end{equation}
so Equation (\ref{SecondConditionForTheHigherOrderSolidHarmonic:IntermediateEquation1}) yields
\begin{equation}
    \label{SecondConditionForTheHigherOrderSolidHarmonic:IntermediateEquation3}
    \nabla_\ulength^2\ulength^2\Pop p_l=2(2l+1)\Pop p_l
\end{equation}
Equation (\ref{SecondConditionForTheHigherOrderSolidHarmonic}) is a direct conclusion of Equation (\ref{SecondConditionForTheHigherOrderSolidHarmonic:IntermediateEquation2}) and (\ref{SecondConditionForTheHigherOrderSolidHarmonic:IntermediateEquation3}).

\section{Proof of Equation (\ref{DirectIntermediateIdentity})}\label{Appendix:ProofOfDirectIntermediateIdentity}
The objective here is to show that, with $p_n\in\Hsp_{n}(\Rsp)$
\begin{equation}
    \label{Appendeix:DirectIntermediateIdentity}
    \Pop^{n+1}\left[(2n+1)\Qop-\ulength^2\Pop\right]p_n=(n+1)^2\nabla^2\Pop^np_n
\end{equation}
Reference \cite{Ackroyd1999:IsotropicSHPn} did this by proving
\begin{equation}
    \label{Appendeix:DirectIntermediateIdentitySplitEquation1}
    \Pop^{n+1}\Qop p_n=(n+1)\nabla^2\Pop^np_n
\end{equation}
and
\begin{equation}
    \label{Appendeix:DirectIntermediateIdentitySplitEquation2}
    \Pop^{n+1}\ulength^2\Pop p_n=n(n+1)\nabla^2\Pop^np_n
\end{equation}
Both equations above can be proved using mathematical induction.

When $n=0$, the two equations become
\begin{equation}
    \Pop\Qop p_0=\nabla^2p_0
\end{equation}
\begin{equation}
    \Pop\ulength^2\Pop p_0=0
\end{equation}
both of which are quite apparent.

Then, assuming that Equation (\ref{Appendeix:DirectIntermediateIdentitySplitEquation1}) and (\ref{Appendeix:DirectIntermediateIdentitySplitEquation2}) are valid for $n=k$, the corresponding equations for $k+1$ are
\begin{equation}
    \label{Appendeix:DirectIntermediateIdentitySplitEquation1Induction}
    \Pop^{k+2}\Qop p_{k+1}=(k+2)\nabla^2\Pop^{k+1}p_{k+1}
\end{equation}
and
\begin{equation}
    \label{Appendeix:DirectIntermediateIdentitySplitEquation2Induction}
    \Pop^{k+2}\ulength^2\Pop p_{k+1}=(k+1)(k+2)\nabla^2\Pop^{k+1}p_{k+1}
\end{equation}
We will tackle the first equation first. By Equation (\ref{IntermediateIdentity(1)})
\begin{equation}
    \Pop^{k+2}\Qop p_{k+1}=\Pop^{k+1}(\Qop\Pop+\nabla^2)p_{k+1}
\end{equation}
By the equation for $n=k$
\begin{equation}
    \Pop^{k+2}\Qop p_{k+1}=\left((k+1)\nabla^2\Pop^{k+1}+\Pop^{k+1}\nabla^2\right)p_{k+1}=(k+2)\nabla^2\Pop^{k+1}p_{k+1}
\end{equation}
So we have reached Equation (\ref{Appendeix:DirectIntermediateIdentitySplitEquation1}), and will move on to prove Equation (\ref{Appendeix:DirectIntermediateIdentitySplitEquation2Induction}). By Equation (\ref{IntermediateIdentity(2)})
\begin{equation}
    \Pop^{k+2}\ulength^2\Pop p_{k+1}=\Pop^{k+1}(\ulength^2\Pop+2\Qop)\Pop p_{k+1}
\end{equation}
By Equation (\ref{Appendeix:DirectIntermediateIdentitySplitEquation1}) and Equation (\ref{Appendeix:DirectIntermediateIdentitySplitEquation2}) for $n=k$
\begin{equation}
    \Pop^{k+2}\ulength^2\Pop p_{k+1}=k(k+1)\nabla^2\Pop^{k+1}p_{k+1}+2(k+1)\nabla^2\Pop^{k+1}p_{k+1}
\end{equation}
So we have also arrived at Equation (\ref{Appendeix:DirectIntermediateIdentitySplitEquation2}) and Equation (\ref{Appendeix:DirectIntermediateIdentity}) is a direct conclusion following Equation (\ref{Appendeix:DirectIntermediateIdentitySplitEquation1}) and (\ref{Appendeix:DirectIntermediateIdentitySplitEquation2}).

\end{appendix}

\end{document}